\documentclass[amsfonts,aps,preprint,nofootinbib]{revtex4-1}
\usepackage{hyperref}
\usepackage{graphicx}
\usepackage[dvips]{epsfig}
\usepackage{latexsym}
\usepackage{amsmath}
\usepackage[english]{babel}
\usepackage{enumerate}
\newcommand{\be}{\begin{equation}}
\newcommand{\ee}{\end{equation}}
\newcommand{\ba}{\begin{eqnarray}}
\newcommand{\ea}{\end{eqnarray}}

\def\a{\alpha}
\def\ah{\hat\a}
\def\ab{\overline{\alpha}}
\def\au{\underline{a}}
\def\b{\beta}
\def\bb{\overline\b}
\def\bh{\hat\b}
\def\bu{\underline{b}}
\def\cu{\underline{c}}
\def\d{\delta}

\def\du{\underline{d}}
\def\dh{\hat\d}
\def\dt{\widetilde d}

\def\e{\epsilon}

\def\eu{\underline{e}}

\def\fu{\underline{f}}
\def\g{\gamma}
\def\gb{\overline\g}
\def\gu{\underline{g}}
\def\gh{\hat\g}
\def\hu{\underline{h}}

\def\l{\lambda}
\def\lt{\widetilde\l}
\def\lh{\widehat\lambda}

\def\o{\omega}
\def\ot{\widetilde\o}
\def\oh{\widehat\o}

\def\O{\Omega}

\def\Ot{\widetilde\O}

\def\t{\theta}

\def\G{\Gamma}

\def\Nh{\hat N}
\def\N{\nabla}
\def\Nb{\overline\nabla}

\def\p{\partial}
\def\pb{\overline\partial}
\def\Jb{\overline J}
\def\Ct{\widetilde C}

\begin{document}

\title{Quantum corrections to $AdS_5\times S^5$Ê left-invariant superstring current algebra}
\author{D\'afni F. Z. Marchioro}
\email{dafnimarchioro@unipampa.edu.br}
\author{Daniel Luiz Nedel}
\email{daniel.nedel@unipampa.edu.br}
\affiliation{Universidade Federal do Pampa, Campus Bag\'e, Travessa 45, n\'umero 1650 - Bairro Malafaia, CEP: 96413-170, Bag\'e, RS, Brasil}

\begin{abstract}
In this work the pure spinor formulation of the superstring is used to study quantum corrections to the left current OPE algebra of the coset $PSU(2,2|4)/SO(4,1)\times SO(5)$ sigma model, which describes the superstring dynamics in the $AdS_5 \times S^5$ background. In particular,  the one loop corrections to the simple poles of the bosonic currents are computed. Unlike the case of the double poles, we show that the simple poles suffer corrections, which are important since the simple poles contribute to the four point amplitudes. We show that the only contribution to the simple poles comes from the pure spinor Lorentz currents.

\end{abstract}

\maketitle

\section{Introduction}

In view of the structural role AdS superstring plays in the construction of viable theoretic frameworks for understanding the non perturbative regime of Yang-Mills theory, the AdS superstrings and the AdS/CFT correspondence are topics of constant and renewed interest in the string literature. Also, on the string side, the quantization of the string in the $AdS_5\times S^5$ space provides an important example of string quantization in a Ramond-Ramond background and it is interesting by itself.  

 Despite all the results concerning the $AdS_5\times S^5$ superstring, owing to the complicated structure of the worldsheet action, the quantum properties of the string sigma model  still remain elusive. In this direction, the one loop conformal invariance was proved in \cite{BBHZZ,Brenno} and the argument for all loop conformal invariance was presented in \cite{Berkovits:2004xu}. In \cite{Mazzucato:2009fv}, the one loop effective action was computed, where it was shown that the  't Hooft coupling $\lambda$ and the $AdS$ radius  are not renormalized at one loop. This result was confirmed  in \cite{quantumopes}, where the double poles of the left invariant currents were calculated at one loop.   
  
In this work, more ingredients of the quantum properties of the $AdS_{5}\times S^{5}$ string sigma model are presented. A basic set of operators in this two dimensional conformal field theory is composed by the left invariant currents.  Since  these currents are not holomorphic even in the classical limit,  their OPEs cannot be deduced from general arguments and we need to develop a perturbative approach.  Although  these currents are not gauge invariant,  they are invariant under global $PSU(2,2|4)$ transformations and they can be used to construct integrated massless vertex operators; also,  they appear in massive unintegrated vertex operators. In the case of the pure spinor superstring, these OPEs play an important role, since the BRST charge is written in terms of the left invariant currents. In addition, the quantum OPEs are needed to unambiguously define the b anti-ghost in $AdS_5\times S^5$, taking into account normal-ordering corrections \cite{nathanluca}.  Besides  these
practical applications, the knowledge  of this current algebra in the worldsheet may shed light into more general aspects of the theory, such as the apparent quantum integrability \footnote{ The classical integrability of the $AdS_5\times S^5$ sigma model was established in the paper \cite{Bena:2003wd} for Green-Schwarz-Metsaev-Tseytlin action and  in \cite{Vallilo:2003nx} for the pure spinor description.  Using cohomological and algebraic renormalization techniques, Berkovits argued that the sigma model still has an infinite number of conserved charges when quantum effects are taken into account \cite{Berkovits:2004xu}. A detailed study of the transfer matrix of the worldsheet was done in \cite{Mikhailov:2007mr}, where it was shown to be a well defined operator in quantum theory.  The literature on this subject is very large, and we did not attempt to give a list of references. The reader can see a list of references in \cite{Luca, reviewvalentina}.  }. 
The tree level OPEs of these currents were computed in \cite{Puletti:2006vb} (see also \cite{Mikhailov:2007mr}). In \cite{quantumopes} a momentum space perturbative approach was developed to calculate the quantum corrections and it was calculated  the double pole corrections.  Surprisingly, in the bosonic sector  most of the possible one-loop corrections vanishes due to spacetime supersymmetry and there is no correction.  The result obtained in this reference  confirms the effective action result obtained in \cite{Mazzucato:2009fv}, and it serves as further evidence that the relation between the 't Hooft coupling and the $AdS$ radius is not renormalized. However, for the simple pole we do not expect this behavior, because the simple poles contribute for the four point amplitudes. 

 Keeping in mind this whole framework and the motivations mentioned above,   we  use the perturbative techniques developed in \cite{quantumopes} and calculated the one loop simple poles of bosonic left invariant currents of the  $AdS_5\times S^5$  pure spinor superstring and  we show that the pure spinor Lorentz currents play an important role in the result. In fact, we show that the only contribution for the simple poles comes from the pure spinor currents.
We present this paper according to the following outline: in section 2, we present a review of the pure spinor superstring in the $AdS_5\times S^5$ background; section 3 is devoted to present the Feynman rules and dimensional regularization used in our calculations; in section 4 we present our results; the conclusions are summarized in section 5 and section 6 is the appendix.

\section{Review of Pure Spinor $AdS_5\times S^5$ Superstring}

In the Green-Schwarz (GS) superstring,  the target space supersymmetries are manifest and  the superspace coordinates are treated more symmetrically with respect to the Ramond-Neveu-Schwarz (RNS) formalism. So, relative to RNS formalism, the GS formalism is more appropriate to study curved  backgrounds  with Ramond-Ramond (RR) fluxes.  Using the GS superstring, Metsaev and Tseytlin constructed the worldsheet action for the type IIB superstring
in $AdS_5\times S^5$  from a geometrical point of view based on a super coset approach \cite{MetsaevTseytlin}.  They showed that the
superstring in $AdS_5 \times S^5$ background can be described using
some currents defined in the superalgebra $psu(2,2|4)$. Those
currents are described in terms of the supervielbein $E_M{}^A$  and 
 are defined in a left-invariant way by
 
 \begin{eqnarray}
  J^A &=& (g^{-1} \p g)^A = \p Z^M E_{M}^A  \nonumber \\
   \Jb^A &=& (g^{-1} \pb g)^A = \pb Z^M E_{M}^A,
   \end{eqnarray}
where $Z^M$ are the curved superspace coordinates and $g$ an element in the coset supergroup
$PSU(2,2|4)/SO(4,1)\times SO(5)$.  The index $A$ denotes $(\au, \a, \ah,
)$ and $a = 0,{\ldots} 4$ for $AdS_5$, $a' = 5,{\ldots} 9$ for $S^5$, $\a = 1,{\ldots}
16$, $\ah = 1,{\ldots} 16$ and $\au$ denotes both $a$ and $a'$.

In spite of the target space manifest supersymmetry of the GS formalism, we encounter, already in flat space, serious difficulties once we try to covariantly quantize the GS superstring. This is because the first and second class constraints that appear in this formalism cannot be separated in a covariant way. A manifest supersymmetric formalism, which can be quantized in a covariant way, was proposed by Berkovits in \cite{BerkovitsCQS}. In this formalism, in order to have standard fermionic kinetic terms, certain ghost fields have to be introduced, satisfying pure spinor constraints. The non-physical degrees of freedom introduced in the theory in this way are later removed through a BRST-like operator $Q$. Although the BRST operator used in pure spinor formalism is not originally constructed in a traditional way, the formalism has passed for all tests and nowadays is a powerful tool to understand superstring theory \footnote{Actually,  in ref. \cite{newBRST}a new interpretation of the pure spinor BRST charge was proposed, which elegantly explains its
origin.}.

In a curved background, the pure spinor sigma model action for the
type II superstring is obtained by adding to the flat action the integrated 
vertex operator for supergravity
massless states and then covariantizing with respect to ten dimensional
$N = 2$ super-reparameterization invariance. The result of doing
this is

\be\label{action} S = {1\over{2\pi\a'}} \int d^2 z ~ ({1\over 2} \p Z^M \pb Z^N
(G_{NM} + B_{NM}) + d_\a  \pb Z^M E_{M}^\a  + \dt_{\ah}
\p Z^M E_{M}^{\ah} + \l^\a \o_\b \pb Z^M \O_{M\a}{}^\b  
\ee $ + \lt^{\ah}
\ot_{\bh}\p Z^M \O_{M\ah}{}^{\bh} + d_\a \dt_{\bh} P^{\a\bh} + \l^\a \o_\b
\dt_{\gh} C_\a{}^{\b\gh} + \lt^{\ah} \ot_{\bh} d_\g
\Ct_{\ah}{}^{\bh\g} + \l^\a \o_\b  \lt^{\ah} \ot_{\bh}
S_{\a\ah}{}^{\b\bh} )+ S_{pure} + S_{FT},$

\noindent where  $B_{NM}$ is the
super two-form potential, and $S_{pure}$ is the
action for the pure spinor ghosts and is the same as in the flat
space case. The pure spinor condition means that they
satisfy $\l^\a \g^{c} _{\a\b} \l^\b =0$ and $\lh^{\ah} \g^{c} _{\ah\bh}
\lh^{\bh} =0$, where $c = 0,{\ldots} 9$ is a tangent space bosonic index. The gravitini and the dilatini
fields  are described by the lowest $\t$-components of the
superfields $C_{\a}{}^{\b\gb}$ and $\Ct_{\ab}{}^{\bb\g}$, while the
Ramond-Ramond field strengths are in the superfield $P^{\a\bb}$ \cite{BerkovitsHowe}. The
dilaton is the theta independent part of the superfield $\Phi$ which
defines the Fradkin-Tseytlin term

\be
S_{FT} = {1\over{2\pi}} \int d^2z  \: r \: \Phi,
\label{ft}
\ee

\noindent where $r$ is the worldsheet curvature. 

In $AdS_5\times S^5$ the non zero component of two form $B_{AB}$,  the five-form Ramond-Ramond field-strength and the curvature are respectively:
\begin{eqnarray}
B_{\a\bh} &=& {1\over 2} (N g_s)^{1\over 4} \sqrt{\a'} \d_{\a\bh} \nonumber \\
P^{\a\bh} &=& {\d^{\a\bh} \over {\sqrt{\a'}(N g_s)^{1\over4}}} \nonumber \\
R_{abcd} &=& -{1\over R^2} \eta_{a[c}\eta_{d]b} ,  \,\,\, 
R_{a'b'c'd'} = {1\over R^2} \eta_{a'[c'}\eta_{d']b'},
\end{eqnarray}
  where $R$  is the radius of $AdS_5$ and $S^5$. As shown in \cite{BerkovitsHowe}, the term containing $S_{\a\ah}^{\b\bh}$ is related to the
constant space-time curvature, while the values of the superfields $C_\a ^{\b\gh}$ and
$\Ct_{\ah}^{\bh\g}$ are zero, as well as
$\O_M ^{(s)}$ and $\Ot _M ^{(s)}$ because they are related to
derivatives of the superfield containing the dilaton, which is
constant for this background. Finally, the terms containing the spin connections will lead to 

\be\label{spinconnections}\l^\a \o_\b \pb Z^M \O_{M\a}{}^\b = 
N_{\au\bu} \Jb^{\au\bu} , \,\, \lh^{\ah} \oh_{\bh} \p Z^M
\Ot_{M\ah}{}^{\bh} = \hat N_{\au\bu} J^{\au\bu} , 
\ee
where $J^{\au\bu} = {1\over 2} \p Z^M \Ot_{M\au\bu}$, $\Jb^{\au\bu} = {1\over 2}
\pb Z^M \O_{M\au\bu}$ and $N^{\au \bu} = {1\over 2}
(\l \g^{\au \bu} \o)$,  $\Nh^{\au \bu} = {1\over 2}
(\lh \g^{\au \bu} \oh)$ are the pure spinors Lorentz currents, which play an important role in pure spinor formalism \footnote{These currents are the ghost part of a redefined Lorentz current $M^{mn}$, necessary for the implementation of a formalism which has manifest supersymmetry and can be covariantly quantized; see \cite{BerkovitsCQS}.}. Using the definition of the currents in terms of the supervielbein  and taking the $AdS_5\times S^5$ values for  the metric $G_{MN}$, the two-form $B_{MN}$ and the Ramond-Ramond flux, after using the equations of motion for $d_{\alpha}$ and scaling the fields (see ref. \cite{quantumopes} for details), the pure spinor superstring in $AdS_5\times S^5$ is written in terms of the $psu(2,2|4)$ currents as \cite{BerkovitsCQS}:

\be\label{adsaction}S = {1\over {2\pi\a'\a^2}} \int d^2 z ({1\over 2 } J^{\au} \Jb ^{\bu}
\eta_{\au \bu} + \d_{\a\bh} (J^\a \Jb^{\bh} - 3 J^{\bh} \Jb ^\a) \ee $$ +
N_{\au \bu}
\Jb^{[\au \bu]} + \Nh_{\au \bu}J^{[\au \bu]} - N_{ab} \Nh^{ab} +
N_{a'b'} \Nh^{a'b'} ) +
S_\l + S_{\lh},
$$
Note also that in (\ref{adsaction}) all
$J$'s, $\Jb$'s and pure spinor Lorentz currents have engineering
dimension one. So, by choosing units in which $2\pi \a' =1$, the action
is given in terms of dimensionless worldsheet fields.

The currents  $(J^A , \Jb^A)$ satisfy
the Maurer-Cartan identities $\p \Jb^A - \pb J^A + [J,\Jb]^A = 0$, so
by making a variation of the action and using those identities, we can
find the equations of motion 
\be
\label{eom}\N \Jb_2 = -[J_1 , \Jb_1] + {1\over 2} [N,\Jb_2] - {1\over
2}[J_2 , \bar N], 
\ee
\be \label{eomJ2}
\bar\N J_2 = [J_3 , \Jb_3] - {1\over 2} [J_2,\bar N] + {1\over
2}[N , \Jb_2] 
\ee
\be
\N \Jb_1 = {1\over 2} [N,\Jb_1]  - {1\over 2} [J_1 , \bar N],
\ee
\be
\Nb J_1 = [J_2 , \Jb_3] + [J_3, \Jb_2] + {1\over 2} [N, \Jb_1] -
{1\over 2} [J_1 , \bar N]
\ee
\be
\Nb J_3 = {1\over 2} [N,\Jb_3]  - {1\over 2} [J_3 , \bar N],
\ee
\be
\N \Jb_3 = -[J_2 , \Jb_1] - [J_1, \Jb_2] + {1\over 2} [N, \Jb_3] -
{1\over 2} [J_3 , \bar N].
\ee
The pure spinors have
also equations of motion, given by $\Nb N = {1\over 2} [N,\hat N]$ and $\N
\hat N = {1\over 2} [N,\hat N]$, where $\N = \p + [J_0 ,\, ]$ and $\bar\N = \pb + [\Jb_0 ,\, ]$. We have 
suppressed the index $A$ and introduced a sub-index
$0,1,2,3$ for the currents. Using this notation, the current can be written in terms of the generators of
$psu(2,2|4)$ as follows:

\begin{eqnarray}
J_0 &=& J^{[\au \bu]}
M_{\au \bu}\nonumber \\
J_1 &=& J^\a Q_\a \nonumber \\
J_2 &=& J^{\au} P_{\au} \nonumber \\
J_3 &=& J^{\ah} {\hat Q}_{\ah} \\
\end {eqnarray}

 \noindent and similarly for the $\Jb$ currents. The structure constants different from zero from the $psu(2,2|4)$ algebra are
 
 \be\label{structureconstants}f^{\cu} _{\a\b} = 2 \g^{\cu} _{\a\b} , \,\,\,\, f^{\cu}
_{\ah\bh} = 2 \g^{\cu} _{\ah\bh} \ee
$$
f^{[ef]}_{\a\bh} = (\g^{ef})_\a {}^\g \d_{\g\bh} = - (\g^{ef})_{\bh}
{}^{\gh} \d_{\a\gh} = f^{[ef]}_{\bh\a},
\,\,\, f^{[e'f']}_{\a\bh} = -(\g^{e'f'})_\a {}^\g
\d_{\g\bh}  = (\g^{e'f'})_{\bh} {}^{\gh}
\d_{\a\gh}  = f^{[e'f']}_{\bh\a}, $$ $$
f^{\bh} _{\a \cu} = - f^{\bh} _{\cu\a} = {1\over 2} (\g_{\cu})_{\a\b} \d^{\b\bh} ,
\,\,\, f^\b _{\ah \cu} = - f^\b _{\cu \ah} = -{1\over 2} (\g_{\cu})_{\ah\bh}
\d^{\b\bh},$$ $$
f_{cd}^{[ef]} = {1\over 2} \d_c ^{[e} \d_d ^{f]} , \,\,\,
f_{c'd'}^{[e'f']} = -{1\over 2} \d_{c'} ^{[e'} \d_{d'} ^{f']}, $$ $$
f_{[\cu \du][\eu \fu]}^{[\gu \hu]} = {1\over 2} (\eta_{\cu \eu} \d_{\du} ^{[\gu} \d_{\fu} ^{\hu]} -
\eta_{\cu \fu} \d_{\du} ^{[\gu} \d_{\eu} ^{\hu]}+\eta_{\du \fu} \d_{\cu} ^{[\gu} \d_{\eu} ^{\hu]} -
\eta_{\du \eu} \d_{\cu} ^{[\gu} \d_{\fu} ^{\hu]}),$$ $$
f^{\fu} _{[\cu \du] \eu} = - f^{\fu} _{\eu [\cu \du]} = \eta_{\eu[\cu}\d_{\du]}^{\fu} , \,\, f^\b
_{[\cu \du]\a} = -f^\b _{\a [\cu \du]} = {1\over 2} (\g_{\cu \du})_\a {}^\b , \,\,
f^{\bh}
_{[\cu \du]\ah} = -f^{\bh} _{\ah [\cu \du]} = {1\over 2} (\g_{\cu \du})_{\ah} {}^{\bh} .
$$

\noindent This $Z_4$ grading for the superalgebra was noted in \cite{BBHZZ}.

\section{Feynmam Rules and Dimensional Regularization}
Now we are going to show the strategy of the  calculations we intend to do. In order to calculate the simple poles corrections to the bosonic part of the algebra,  we need to calculate contributions to the expectation value
 $\left\langle J^{\au}(y)J^{\bu}(z)\right\rangle$
  perturbatively, including double contractions (one  loop) with  contributions of  one classical field.  In order to do this, we perform a background field expansion as in \cite{BBHZZ}, 
\cite{Brenno} and \cite{BedoyaYM}, choosing  a classical background
given by an element $g_0$ in the supergroup and parametrizing the
quantum fluctuations by $X$ as $g = g_0 e^{\a X}$. Then, the currents
can be written as 
\be\label{currentsplit}
J = g^{-1}\p g = e^{-\a X}J_0 e^{\a X} +  e^{-\a
    X}\p e^{\a X}, \ee
    $$ 
\Jb = g^{-1}\pb g = e^{-\a X}\Jb_0 e^{\a X} +  e^{-\a
    X}\pb e^{\a X}. $$
    By expanding the exponentials in (\ref{currentsplit}) we get:

\be\label{Jsplit}J = J_0 + \a (\p X + [J_0 , X]) + {{\a^2}\over 2}([\p X ,
  X] + [[J_0,X],X]) + {{\a^3}\over {3!}}([[\p X , X],X]+[[[J_0
  ,X],X],X]) + ... , 
  \ee
where $J_0$ denotes the
classical part of $J$ and we have a  similar expression for  $\Jb$. The commutators can be evaluated  using
the structure constants of the $psu(2,2|4)$ Lie superalgebra given by
(\ref{structureconstants}). Here we are going to use the $SO(1,4)\times SO(5)$ gauge invariance to fix $X_0=0$. In addition to expansion of the matter part of the currents, we need to expand the ghost part as follow:

\be\label{pslcexp}N_{\au\bu} = N_{\au\bu}^{(0)} + \a N^{(1)}_{\au\bu} +
\a^2 N^{(2)}_{\au\bu}
\ee
and similarly for $\hat N_{\au\bu}$. Now, the pure spinor Lorentz
currents have the following behavior

\be\label{NoneNone}N_{\au\bu}^{(1)} (y) N_{\cu\du}^{(1)}(z) \to 
{{\eta_{\cu[\bu} N_{\au]\du}^{(0)}(z) - \eta_{\du[\bu}
N_{\au]\cu}^{(0)}(z)}\over{y-z}}, 
\ee

\be\label{NtwoNtwo}N_{\au\bu}^{(2)} (y) N_{\cu\du}^{(2)}(z) \to
-3{{\eta_{\au[\du}\eta_{\cu]\bu}}\over{(y-z)^2}}. 
\ee

 Replacing the expansions  of the currents  in (\ref{adsaction}), one can identify the kinetic 
piece $S_p$ of the action 
\be\label{Sp}S_p = \int d^2 z \left({1\over 2} \p X^{\au} \pb X^{\bu}
  \eta_{\au\bu} + 4 \d_{\a\bh}\p X^\a \pb X^{\bh} \right),
  \ee
from which we obtain the propagators in coordinate space 

\begin{eqnarray}
X^a (y) X^b (z) \rightarrow -\eta^{ab} \ln |y-z|^2 , \:\:\:\:
X^{a'} (y) X^{b'} (z) \rightarrow -\delta^{a'b'} \ln |y-z|^2 \nonumber\\
X^{\alpha} (y) X^{\hat\beta} (z) \rightarrow -\frac{1}{4}
\delta^{\alpha\hat\beta} \ln |y-z|^2 .
\label{propb}
\end{eqnarray}

The reminder terms of the background expansion will provide the
vertices of the theory. In order to calculate the one loop simple pole corrections, we need the expansion
of the action up to three terms in the quantum fields and one classical field. There are a lot of terms in the expansion of the action. Let us just write the terms that will contribute to the expectation value
 $\left\langle J^{\au}(y)J^{\bu}(z)\right\rangle$. The  terms with one classical field and two quantum fields are:
\begin{eqnarray} 
 S(X^2 , J) = \int d^2 z \left[ \eta_{\underline{ab}} \eta_{\underline{cd}} \partial X^{\au} X^{\underline{c}} {\bar J}^{[\underline{db}]} + \eta_{\underline{ab}} \eta_{\underline{cd}} {\bar\partial} X^{\au} X^{\underline{c}} J^{[\underline{db}]} + \frac{1}{2} \partial X^{\alpha} X^{\hat\gamma} {\bar J}^{[\underline{cd}]} (\gamma_{\underline{cd}})_{\hat\gamma}^{\:\:\:\hat\beta} \delta_{\alpha\hat\beta} \right. \nonumber\\
 \left. + \frac{3}{2} {\bar\partial} X^{\alpha} X^{\hat\gamma} J^{[\underline{cd}]} (\gamma_{\underline{cd}})_{\hat\gamma}^{\:\:\:\hat\beta} \delta_{\alpha\hat\beta} -\frac{3}{2} \partial X^{\hat\beta} X^{\gamma} {\bar J}^{[\underline{cd}]} (\gamma_{\underline{cd}})_{\gamma}^{\:\:\:\alpha} \delta_{\alpha\hat\beta} - \frac{1}{2} {\bar\partial} X^{\hat\beta} X^{\gamma} J^{[\underline{cd}]} (\gamma_{\underline{cd}})_{\gamma}^{\:\:\:\alpha} \delta_{\alpha\hat\beta}\right] 
 \label{s2j}
 \end{eqnarray}
 
 \noindent and
 
 \begin{eqnarray}
 S(X^2 , N) = \int d^2 z \left[ \frac{1}{2} {\bar\partial} X^{c} X^{d} N_{cd} - \frac{1}{2} {\bar\partial} X^{c'} X^{d'} N_{c'd'} + \frac{1}{2} {\partial} X^{c} X^{d} {\hat N}_{cd} - \frac{1}{2} {\partial} X^{c'} X^{d'} {\hat N}_{c'd'} \right. \nonumber\\
 \left. -\frac{1}{2} {\bar\partial} X^{\alpha} X^{\hat\beta} (\gamma^{ab})_{\alpha}^{\:\:\:\gamma} \delta_{\gamma\hat\beta} N_{ab} + \frac{1}{2} {\bar\partial} X^{\alpha} X^{\hat\beta} (\gamma^{a'b'})_{\alpha}^{\:\:\:\gamma} \delta_{\gamma\hat\beta} N_{a'b'} - \frac{1}{2} {\bar\partial} X^{\hat\beta} X^{\alpha} (\gamma^{ab})_{\alpha}^{\:\:\:\gamma} \delta_{\gamma\hat\beta} N_{ab}\right. \nonumber\\
 \left. + \frac{1}{2} {\bar\partial} X^{\hat\beta} X^{\alpha} (\gamma^{a'b'})_{\alpha}^{\:\:\:\gamma} \delta_{\gamma\hat\beta} N_{a'b'} -\frac{1}{2} {\partial} X^{\alpha} X^{\hat\beta} (\gamma^{ab})_{\alpha}^{\:\:\:\gamma} \delta_{\gamma\hat\beta} {\hat N}_{ab} + \frac{1}{2} {\partial} X^{\alpha} X^{\hat\beta} (\gamma^{a'b'})_{\alpha}^{\:\:\:\gamma} \delta_{\gamma\hat\beta} {\hat N}_{a'b'}  \right. \nonumber\\
\left.  - \frac{1}{2} {\partial} X^{\hat\beta} X^{\alpha} (\gamma^{ab})_{\alpha}^{\:\:\:\gamma} \delta_{\gamma\hat\beta} {\hat N}_{ab} + \frac{1}{2} {\partial} X^{\hat\beta} X^{\alpha} (\gamma^{a'b'})_{\alpha}^{\:\:\:\gamma} \delta_{\gamma\hat\beta} {\hat N}_{a'b'}\right].
\label{sxxn}
 \end{eqnarray}
 The terms with three quantum fields  are
 
\begin{eqnarray}
\label{SXXX}
S(X^3) &=& {\a \over 4} \int d^2 z[ \p X^{\au} \pb X^\a X^\b
(\g_{\au})_{\a\b} -  \p X^{\au} \pb X^{\ah} X^{\bh}
(\g_{\au})_{\ah\bh} -  \pb X^{\au} \p X^\a X^\b
(\g_{\au})_{\a\b} +  \pb X^{\au} \p X^{\ah} X^{\bh}
(\g_{\au})_{\ah\bh} \nonumber\\
 &+&2 X^{\au} \p X^\a \pb X^\b (\g_{\au})_{\a\b}
- 2 X^{\au} \p X^{\ah} \pb X^{\bh} (\g_{\au})_{\ah\bh} ].
\end{eqnarray}
Integrating by parts the first line we obtain
\be\label{SXXXa}S(X^3) = \a \int d^2 z  [ X^{\au} \p X^\a \pb X^\b
(\g_{\au})_{\a\b} -   X^{\au} \p X^{\ah} \pb X^{\bh}
(\g_{\au})_{\ah\bh} ].
\ee 
 The terms with three quantum fields and one classical field are
 \begin{eqnarray}
 S(X^3 , J)= \alpha \int d^2 z \left[-\frac{1}{2} \partial X^{\alpha} X^{\beta} X^{\cu} \eta_{\cu\du} {\bar J}^{[\du\bu]} (\gamma_{\bu})_{\alpha\beta} + \frac{1}{4} \partial X^{\delta} X^{\gamma} X^{\cu} {\bar J}^{[\au\bu]} (\gamma_{\au\bu} \gamma_{\cu})_{\gamma\delta} \right. \nonumber\\
\left. +\frac{1}{2} {\bar\partial} X^{\alpha} X^{\beta} X^{\cu} \eta_{\cu\du} J^{[\du\bu]} (\gamma_{\bu})_{\alpha\beta} - \frac{1}{4} {\bar\partial} X^{\delta} X^{\gamma} X^{\cu} J^{[\au\bu]} (\gamma_{\au\bu} \gamma_{\cu})_{\gamma\delta} + \frac{1}{8} \partial X^{\cu} X^{\delta} X^{\gamma} {\bar J}^{[\au\bu]} (\gamma_{\au\bu} \gamma_{\cu})_{\delta\gamma} \right. \nonumber\\
\left. -\frac{1}{8} {\bar\partial} X^{\cu} X^{\delta} X^{\gamma} J^{[\au\bu]} (\gamma_{\au\bu} \gamma_{\cu})_{\delta\gamma} +\frac{1}{2} \partial X^{\hat\alpha} X^{\hat\beta} X^{\cu} \eta_{\cu\du} {\bar J}^{[\du\bu]} (\gamma_{\bu})_{\hat\alpha\hat\beta} - \frac{1}{4} \partial X^{\hat\delta} X^{\hat\gamma} X^{\cu} {\bar J}^{[\au\bu]} (\gamma_{\au\bu} \gamma_{\cu})_{\hat\gamma\hat\delta} \right. \nonumber\\
\left. -\frac{1}{2} {\bar\partial} X^{\hat\alpha} X^{\hat\beta} X^{\cu} \eta_{\cu\du} J^{[\du\bu]} (\gamma_{\bu})_{\hat\alpha\hat\beta} + \frac{1}{4} {\bar\partial} X^{\hat\delta} X^{\hat\gamma} X^{\cu} J^{[\au\bu]} (\gamma_{\au\bu} \gamma_{\cu})_{\hat\gamma\hat\delta} - \frac{1}{8} \partial X^{\cu} X^{\hat\delta} X^{\hat\gamma} {\bar J}^{[\au\bu]} (\gamma_{\au\bu} \gamma_{\cu})_{\hat\delta\hat\gamma}\right. \nonumber\\
\left. +\frac{1}{8} {\bar\partial} X^{\cu} X^{\hat\delta} X^{\hat\gamma} J^{[\au\bu]} (\gamma_{\au\bu} \gamma_{\cu})_{\hat\delta\hat\gamma}\right]
 \label{s3j}
 \end{eqnarray}
 
 \noindent and
 
 \begin{eqnarray}
 S(X^3 , N) = \alpha \int d^2 z \left[ \frac{1}{2\times 3!} {\bar\partial} X^{\hat\alpha} X^{\cu} (\gamma_{\cu})_{\hat\alpha\hat\beta} \delta^{\alpha\hat\beta} X^{\hat\gamma} (\gamma^{ab})_{\alpha}^{\:\:\:\gamma} \delta_{\gamma\hat\gamma} N_{ab}  \right.\nonumber\\
 \left. - \frac{1}{2\times 3!} {\bar\partial} X^{\hat\alpha} X^{\cu} (\gamma_{\cu})_{\hat\alpha\hat\beta} \delta^{\alpha\hat\beta} X^{\hat\gamma} (\gamma^{a'b'})_{\alpha}^{\:\:\:\gamma} \delta_{\gamma\hat\gamma} N_{a'b'} 
 - \frac{1}{2\times 3!} X^{\hat\alpha} {\bar\partial} X^{\cu} (\gamma_{\cu})_{\hat\alpha\hat\beta} \delta^{\alpha\hat\beta} X^{\hat\gamma} (\gamma^{ab})_{\alpha}^{\:\:\:\gamma} \delta_{\gamma\hat\gamma} N_{ab} \right.\nonumber\\
\left. + \frac{1}{2\times 3!} X^{\hat\alpha} {\bar\partial} X^{\cu} (\gamma_{\cu})_{\hat\alpha\hat\beta} \delta^{\alpha\hat\beta} X^{\hat\gamma} (\gamma^{a'b'})_{\alpha}^{\:\:\:\gamma} \delta_{\gamma\hat\gamma} N_{a'b'} 
- \frac{1}{2\times 3!} {\bar\partial} X^{\alpha} X^{\cu} (\gamma_{\cu})_{\alpha\beta} \delta^{\beta\hat\beta} X^{\gamma} (\gamma^{ab})_{\gamma}^{\:\:\:\delta} \delta_{\delta\hat\beta} N_{ab} \right. \nonumber\\
\left. + \frac{1}{2\times 3!} {\bar\partial} X^{\alpha} X^{\cu} (\gamma_{\cu})_{\alpha\beta} \delta^{\beta\hat\beta} X^{\gamma} (\gamma^{a'b'})_{\gamma}^{\:\:\:\delta} \delta_{\delta\hat\beta} N_{a'b'} + \frac{1}{2\times 3!}  X^{\alpha} {\bar\partial} X^{\cu} (\gamma_{\cu})_{\alpha\beta} \delta^{\beta\hat\beta} X^{\gamma} (\gamma^{ab})_{\gamma}^{\:\:\:\delta} \delta_{\delta\hat\beta} N_{ab} \right. \nonumber\\
\left. - \frac{1}{2\times 3!}  X^{\alpha} {\bar\partial} X^{\cu} (\gamma_{\cu})_{\alpha\beta} \delta^{\beta\hat\beta} X^{\gamma} (\gamma^{a'b'})_{\gamma}^{\:\:\:\delta} \delta_{\delta\hat\beta} N_{a'b'} -\frac{1}{3!} {\bar\partial} X^{\alpha} X^{\beta} \gamma^{c}_{\alpha\beta} X^{d} \delta^{[a}_{c} \delta^{b]}_{d} N_{ab} \right. \nonumber\\
\left. + \frac{1}{3!} {\bar\partial} X^{\alpha} X^{\beta} \gamma^{c'}_{\alpha\beta} X^{d'} \delta^{[a'}_{c'} \delta^{b']}_{d'} N_{a'b'} - \frac{1}{3!} {\bar\partial} X^{\hat\alpha} X^{\hat\beta} \gamma^{c}_{\hat\alpha\hat\beta} X^{d} \delta^{[a}_{c} \delta^{b]}_{d} N_{ab} + \frac{1}{3!} {\bar\partial} X^{\hat\alpha} X^{\hat\beta} \gamma^{c'}_{\hat\alpha\hat\beta} X^{d'} \delta^{[a'}_{c'} \delta^{b']}_{d'} N_{a'b'}\right. \nonumber\\
\left. + \frac{1}{2\times 3!} {\partial} X^{\hat\alpha} X^{\cu} (\gamma_{\cu})_{\hat\alpha\hat\beta} \delta^{\alpha\hat\beta} X^{\hat\gamma} (\gamma^{ab})_{\alpha}^{\:\:\:\gamma} \delta_{\gamma\hat\gamma} {\hat N}_{ab}  - \frac{1}{2\times 3!} {\partial} X^{\hat\alpha} X^{\cu} (\gamma_{\cu})_{\hat\alpha\hat\beta} \delta^{\alpha\hat\beta} X^{\hat\gamma} (\gamma^{a'b'})_{\alpha}^{\:\:\:\gamma} \delta_{\gamma\hat\gamma} {\hat N}_{a'b'} \right. \nonumber\\
\left. - \frac{1}{2\times 3!} X^{\hat\alpha} {\partial} X^{\cu} (\gamma_{\cu})_{\hat\alpha\hat\beta} \delta^{\alpha\hat\beta} X^{\hat\gamma} (\gamma^{ab})_{\alpha}^{\:\:\:\gamma} \delta_{\gamma\hat\gamma} {\hat N}_{ab} 
+ \frac{1}{2\times 3!} X^{\hat\alpha} {\partial} X^{\cu} (\gamma_{\cu})_{\hat\alpha\hat\beta} \delta^{\alpha\hat\beta} X^{\hat\gamma} (\gamma^{a'b'})_{\alpha}^{\:\:\:\gamma} \delta_{\gamma\hat\gamma} {\hat N}_{a'b'} \right. \nonumber\\
\left. - \frac{1}{2\times 3!} {\partial} X^{\alpha} X^{\cu} (\gamma_{\cu})_{\alpha\beta} \delta^{\beta\hat\beta} X^{\gamma} (\gamma^{ab})_{\gamma}^{\:\:\:\delta} \delta_{\delta\hat\beta} {\hat N}_{ab} 
+ \frac{1}{2\times 3!} {\partial} X^{\alpha} X^{\cu} (\gamma_{\cu})_{\alpha\beta} \delta^{\beta\hat\beta} X^{\gamma} (\gamma^{a'b'})_{\gamma}^{\:\:\:\delta} \delta_{\delta\hat\beta} {\hat N}_{a'b'} \right. \nonumber\\
\left. + \frac{1}{2\times 3!}  X^{\alpha} {\partial} X^{\cu} (\gamma_{\cu})_{\alpha\beta} \delta^{\beta\hat\beta} X^{\gamma} (\gamma^{ab})_{\gamma}^{\:\:\:\delta} \delta_{\delta\hat\beta} {\hat N}_{ab} 
- \frac{1}{2\times 3!}  X^{\alpha} {\partial} X^{\cu} (\gamma_{\cu})_{\alpha\beta} \delta^{\beta\hat\beta} X^{\gamma} (\gamma^{a'b'})_{\gamma}^{\:\:\:\delta} \delta_{\delta\hat\beta} {\hat N}_{a'b'} \right. \nonumber\\
\left. -\frac{1}{3!} {\partial} X^{\alpha} X^{\beta} \gamma^{c}_{\alpha\beta} X^{d} \delta^{[a}_{c} \delta^{b]}_{d} {\hat N}_{ab} 
+ \frac{1}{3!} {\partial} X^{\alpha} X^{\beta} \gamma^{c'}_{\alpha\beta} X^{d'} \delta^{[a'}_{c'} \delta^{b']}_{d'} {\hat N}_{a'b'}  - \frac{1}{3!} {\partial} X^{\hat\alpha} X^{\hat\beta} \gamma^{c}_{\hat\alpha\hat\beta} X^{d} \delta^{[a}_{c} \delta^{b]}_{d} {\hat N}_{ab} \right.\nonumber\\ 
\left. + \frac{1}{3!} {\partial} X^{\hat\alpha} X^{\hat\beta} \gamma^{c'}_{\hat\alpha\hat\beta} X^{d'} \delta^{[a'}_{c'} \delta^{b']}_{d'} {\hat N}_{a'b'}\right] .\nonumber\\
\end{eqnarray}
Finally, we need terms with quantum pure spinor Lorentz
currents. These terms will contribute to diagram 21 and they are:
\begin{equation}
 S(X^2 , N^{(1)}) = \int d^2 z \left[ \frac{1}{2} {\bar\partial} X^{c} X^{d} N^{(1)}_{cd} - \frac{1}{2} {\bar\partial} X^{c'} X^{d'} N^{(1)}_{c'd'} + \frac{1}{2} {\partial} X^{c} X^{d} {\hat N}^{(1)}_{cd} - \frac{1}{2} {\partial} X^{c'} X^{d'} {\hat N}^{(1)}_{c'd'} \right]
 \label{s3xn} 
 \end{equation}

 Terms with four quantum fields will not contribute since two dimensional tadpoles vanish in the regularization scheme  we are going to use. Given the propagators and the interactions, it is then straightforward to write down
coordinate space expressions for the Feynman rules of the diagrams
that appear in the  expectation value
 $\left\langle J^{\au}(y)J^{\bu}(z)\right\rangle$. However, there are infrared and ultraviolet divergences which 
produce ambiguities in the coordinate space integrals. In order to circumvent this problem, we use  momentum space Feynman rules with a prescription for 
worldsheet dimensional regularization \cite {boer, skenderis, nedel}. After the calculation of the OPE´s in momentum space,  the results are written again in
coordinate space by using an inverse Fourier transformation.

The two dimensional prescription for dimensional regularization consists in keeping all the interactions in exactly two dimensions, but the kinetic terms (and hence the denominators of the propagators) will be in $d=2-2\epsilon$ dimensions. 
  
  We are going to use the definition $d^2 k= \frac{dk_xdk_y}{\pi}$. With this choice there is no $\pi$ dependence in the results and the Green function $G(y,z)$ is represented as
  \begin{equation}
  G(y,z)= \int d^2k\frac{e^{ik(y-z)+i\bar{k}(\bar{y}-\bar{z})}}{\left.k\right.^2}.
  \end{equation}
              
The  momentum space propagators  look like

\be\label{propms}X^{\au} (k) X^{\bu} (l) \rightarrow \eta^{\au \bu}{\d^2
(k+l)\over{|k|^2}}, \,\,\, X^\a (k) X^{\bh} (l) \rightarrow {1\over
4}\d^{\a\bh}{\d^2 (k+l)\over{|k|^2}}.
\ee

To work out the corresponding expression for the OPE's in  momentum space, we use the dimensional regularization prescription and include a factor $\Gamma(1-\epsilon)(4\pi)^{-\epsilon}(2\pi)^{2\epsilon}$ for each loop; this will remove the Euler constant (the G-scheme \cite {gesquem}).
All the integrals we need to compute in the momentum space come from
the formula
\begin{eqnarray}
&&\int d^{2}p\frac{
p^{a}\overline{p}^{b}}{[\left| p\right| ^{2}]^{\alpha }[\left|
p-k\right|^2
]^{\beta }} = \nonumber \\
&&k^{a+1-\alpha -\beta }\overline{k}^{b+1-\alpha -\beta }\left| \frac{k^{2}}{
\mu ^{2}}\right| ^{-\epsilon }
 \times \sum_{i=0}^{i=a}\left(
\begin{array}{c}
a \\
i
\end{array}
\right) [\frac{\Gamma \left( 2-\alpha -\beta +b+i-\epsilon \right) }{
\Gamma \left( 2-2\epsilon -\alpha -\beta +i+b\right) }\nonumber \\
&&\times \frac{\Gamma \left( \alpha +\beta -1-i+\epsilon \right)}{\Gamma \left( 1+\epsilon \right) \mu ^{-2\epsilon }} \Gamma \left(
1-\epsilon -\beta +i\right) ] \label{regul},
\end{eqnarray}
where $\mu$ is the usual mass parameter of the dimensional
regularization. 

Since the main part of this paper is rather technical and we want to be as clear as possible, we end this section showing in detail the calculation of one diagram. Restricting the expansion (\ref{Jsplit}) to the case with one classical
current, we can write

\begin{eqnarray}\label{J_0J_0} 
\langle J^{\au} (y) J^{\bu}(z)\rangle = \a^2 \langle
\p X^{\au}(y) \p X^{\bu} (z) \rangle + \alpha^2 \langle \partial X^{\au} (y) X^{\underline{e}} (z) \rangle J^{[\underline{cd}]} \eta_{\underline{e}[\underline{c}}\delta^{\bu}_{\underline{d}}  +\alpha^2 \langle X^{\underline{e}} (y) \partial X^{\bu} (z) \rangle J^{[\underline{cd}]} \eta_{\underline{e}[\underline{c}} \delta^{\au}_{\underline{d}]} \nonumber\\
- \a^3 \langle \p X^{\au} (y) \p
X^\a  X^{\b} (z)\rangle \g^{\bu}_{\a\b} - \a^3 \langle \p X^{\au} (y) \p
X^{\ah}  X^{\bh} (z)\rangle \g^{\bu}_{\ah\bh} 
-\frac{\alpha^3}{2} \langle \partial X^{\au} (y) X^{\gamma} X^{\beta} (z) \rangle J^{[\underline{cd}]} (\gamma_{\underline{cd}})_{\gamma}^{\:\:\:\alpha} \gamma^{\bu}_{\alpha\beta} \nonumber\\
-\frac{\alpha^3}{2} \langle \partial X^{\au} (y) X^{\hat\gamma} X^{\hat\beta} (z) \rangle J^{[\underline{cd}]} (\gamma_{\underline{cd}})_{\hat\gamma}^{\:\:\:\hat\alpha} \gamma^{\bu}_{\hat\alpha\hat\beta} - \a^3 \langle \p X^\a  X^{\b} (y) \p X^{\bu}(z)\rangle \g^{\au}_{\a\b} 
- \a^3 \langle \p X^{\ah}  X^{\bh} (y) \p X^{\bu}(z)\rangle
\g^{\au}_{\ah\bh} \nonumber\\
-\alpha^3 \langle \partial X^{\alpha} X^{\beta} (y) X^{\underline{e}} (z) \rangle J^{[\underline{cd}]} \gamma^{\au}_{\alpha\beta} \eta_{\underline{e}[\underline{c}} \delta^{\bu}_{\underline{d}]} -\alpha^3 \langle \partial X^{\hat\alpha} X^{\hat\beta} (y) X^{\underline{e}} (z) \rangle J^{[\underline{cd}]} \gamma^{\au}_{\hat\alpha\hat\beta} \eta_{\underline{e}[\underline{c}} \delta^{\bu}_{\underline{d}]} \nonumber\\
 - \alpha^3 \langle X^{\underline{e}} (y) \partial X^{\alpha} X^{\beta} (z) \rangle J^{[\underline{cd}]} \eta_{\underline{e}[\underline{c}} \delta^{\au}_{\underline{d}]} \gamma^{\bu}_{\alpha\beta}
 - \alpha^3 \langle X^{\underline{e}} (y) \partial X^{\hat\alpha} X^{\hat\beta} (z) \rangle J^{[\underline{cd}]} \eta_{\underline{e}[\underline{c}} \delta^{\au}_{\underline{d}]} \gamma^{\bu}_{\hat\alpha\hat\beta} \nonumber\\
 -\frac{\alpha^3}{2} \langle X^{\gamma} X^{\beta} (y) \partial X^{\bu} (z) \rangle J^{[\underline{cd}]} (\gamma_{\underline{cd}})_{\gamma}^{\:\:\:\alpha} \gamma^{\au}_{\alpha\beta} -\frac{\alpha^3}{2} \langle X^{\hat\gamma} X^{\hat\beta} (y) \partial X^{\bu} (z) \rangle J^{[\underline{cd}]} (\gamma_{\underline{cd}})_{\hat\gamma}^{\:\:\:\hat\alpha} \gamma^{\au}_{\hat\alpha\hat\beta} \nonumber\\
+ \a^4 \langle \p X^\a X^\b (y) \p X^{\gh} X^{\dh} (z)
\rangle \g^{\au}_{\a\b} \g^{\bu}_{\gh\dh}  + \a^4 \langle \p X^{\ah}
X^{\bh} (y) \p X^{\g} X^{\d} (z)
\rangle \g^{\au}_{\ah\bh} \g^{\bu}_{\g\d} \nonumber\\
+\frac{\alpha^4}{2} \langle \partial X^{\alpha} X^{\beta} (y) X^{\hat\gamma} X^{\hat\delta} (z) \rangle J^{[\underline{cd}]} \gamma^{\au}_{\alpha\beta} (\gamma_{\underline{cd}})_{\hat\gamma}^{\:\:\:\hat\lambda} \gamma^{\bu}_{\hat\lambda\hat\delta} +\frac{\alpha^4}{2} \langle \partial X^{\hat\alpha} X^{\hat\beta} (y) X^{\gamma} X^{\delta} (z) \rangle J^{[\underline{cd}]} \gamma^{\au}_{\hat\alpha\hat\beta} (\gamma_{\underline{cd}})_{\gamma}^{\:\:\:\lambda} \gamma^{\bu}_{\lambda\delta} \nonumber\\
 + \frac{\alpha^4}{2} \langle X^{\hat\gamma} X^{\hat\delta} (y) \partial X^{\alpha} X^{\beta} (z) \rangle J^{[\underline{cd}]} (\gamma_{\underline{cd}})_{\hat\gamma}^{\:\:\:\hat\lambda} \gamma^{\au}_{\hat\lambda\hat\delta} \gamma^{\bu}_{\alpha\beta} 
+ \frac{\alpha^4}{2} \langle X^{\gamma} X^{\delta} (y) \partial X^{\hat\alpha} X^{\hat\beta} (z) \rangle J^{[\underline{cd}]} (\gamma_{\underline{cd}})_{\gamma}^{\:\:\:\lambda} \gamma^{\au}_{\lambda\delta} \gamma^{\bu}_{\hat\alpha\hat\beta}. \nonumber\\
\label{expJJ}
\end{eqnarray}
Let's focus on the first term in (\ref{J_0J_0}). The diagrams that contribute to this expectation value are 4, 5, 6, 7, 8 and 21. Let's calculate the diagram 7.  We can form this one-loop diagram
by using the two terms in the right-hand side of (\ref{SXXXa}), which
will come from the expansion of the exponential of minus the action at
second order. Also we need: the terms $\displaystyle\frac{1}{2} \partial X^{\alpha} X^{\hat\gamma} {\bar J}^{[\cu\du]} (\gamma_{\cu\du})_{\hat\gamma}^{\:\:\:\hat\beta} \delta_{\alpha\hat\beta}$, $-\displaystyle\frac{3}{2} \partial X^{\hat\beta} X^{\gamma} {\bar J}^{[\cu\du]} (\gamma_{\cu\du})_{\gamma}^{\:\:\:\alpha} \delta_{\alpha\hat\beta}$,  $-\displaystyle\frac{1}{2} {\bar\partial} X^{\hat\beta} X^{\gamma} J^{[\cu\du]} (\gamma_{\cu\du})_{\gamma}^{\:\:\:\alpha} \delta_{\alpha\hat\beta}$ and $\displaystyle\frac{3}{2} {\bar\partial} X^{\alpha} X^{\hat\gamma} J^{[\cu\du]} (\gamma_{\cu\du})_{\hat\gamma}^{\:\:\:\hat\beta} \delta_{\alpha\hat\beta}$ of (\ref {s2j}) (contributions of $J^{\au\bu}$ and ${\bar J}^{\au\bu}$); the terms $-\displaystyle\frac{1}{2} {\bar\partial} X^{\alpha} X^{\hat\beta} (\gamma^{ab})_{\alpha}^{\:\:\:\gamma} \delta_{\gamma\hat\beta} N_{ab}$, $\displaystyle\frac{1}{2} {\bar\partial} X^{\alpha} X^{\hat\beta} (\gamma^{a'b'})_{\alpha}^{\:\:\:\gamma} \delta_{\gamma\hat\beta} N_{a'b'}$, $- \displaystyle\frac{1}{2} {\bar\partial} X^{\hat\beta} X^{\alpha} (\gamma^{ab})_{\alpha}^{\:\:\:\gamma} \delta_{\gamma\hat\beta} N_{ab}$, $\displaystyle\frac{1}{2} {\bar\partial} X^{\hat\beta} X^{\alpha} (\gamma^{a'b'})_{\alpha}^{\:\:\:\gamma} \delta_{\gamma\hat\beta} N_{a'b'}$, $-\displaystyle\frac{1}{2} {\partial} X^{\alpha} X^{\hat\beta} (\gamma^{ab})_{\alpha}^{\:\:\:\gamma} \delta_{\gamma\hat\beta} {\hat N}_{ab}$, $\displaystyle\frac{1}{2} {\partial} X^{\alpha} X^{\hat\beta} (\gamma^{a'b'})_{\alpha}^{\:\:\:\gamma} \delta_{\gamma\hat\beta} {\hat N}_{a'b'}$, $- \displaystyle\frac{1}{2} {\partial} X^{\hat\beta} X^{\alpha} (\gamma^{ab})_{\alpha}^{\:\:\:\gamma} \delta_{\gamma\hat\beta} {\hat N}_{ab}$ and $\displaystyle\frac{1}{2} {\partial} X^{\hat\beta} X^{\alpha} (\gamma^{a'b'})_{\alpha}^{\:\:\:\gamma} \delta_{\gamma\hat\beta} {\hat N}_{a'b'}$ of (\ref{sxxn}) (contributions of $N_{\au\bu}$ and ${\hat N}_{\au\bu}$); and the two terms of (\ref{SXXXa}). So, in momentum space, using the contractions (\ref{propms}), the contribution of diagram 7 to $\a^2 \langle
\p X^{\au}(y) \p X^{\bu} (z) \rangle$ is

\begin{eqnarray}
D_7 = -2i\alpha^4 \frac{{\bar J}^{[\au\bu]}}{{\bar k}^2} \int d^2 p \frac{(p^3 ({\bar k} - {\bar p})^2 + {\bar p}^2 p (k-p)^2 -2 p^2 {\bar p} (k-p) ({\bar k} - {\bar p}))}{|p|^4 |k-p|^2} \nonumber\\
- 2i\alpha^4 \frac{J^{[\au\bu]}}{{\bar k}^2} \int d^2 p \frac{({\bar p}^3 ( k - p)^2 + p^2 {\bar p} ({\bar k}- {\bar p})^2 -2 {\bar p}^2 p (k-p) ({\bar k} - {\bar p}))}{|p|^4 |k-p|^2} \nonumber\\
- i\alpha^4 \frac{N^{ab}}{{\bar k}^2} \int d^2 p \frac{({\bar p}^3 ( k - p)^2 + p^2 {\bar p} ({\bar k}- {\bar p})^2 -2 {\bar p}^2 p (k-p) ({\bar k} - {\bar p}))}{|p|^4 |k-p|^2} \nonumber\\
+ i\alpha^4 \frac{N^{a'b'}}{{\bar k}^2} \int d^2 p \frac{({\bar p}^3 ( k - p)^2 + p^2 {\bar p} ({\bar k}- {\bar p})^2 -2 {\bar p}^2 p (k-p) ({\bar k} - {\bar p}))}{|p|^4 |k-p|^2} \nonumber\\
- i\alpha^4 \frac{{\hat N}^{ab}}{{\bar k}^2} \int d^2 p \frac{(p^3 ({\bar k} - {\bar p})^2 + {\bar p}^2 p (k-p)^2 -2 p^2 {\bar p} (k-p) ({\bar k} - {\bar p}))}{|p|^4 |k-p|^2} \nonumber\\
+i\alpha^4 \frac{{\hat N}^{a'b'}}{{\bar k}^2} \int d^2 p \frac{(p^3 ({\bar k} - {\bar p})^2 + {\bar p}^2 p (k-p)^2 -2 p^2 {\bar p} (k-p) ({\bar k} - {\bar p}))}{|p|^4 |k-p|^2}. 
\end{eqnarray}
For this diagram, we have an overall factor of $\displaystyle\frac{1}{3!}$ coming from the expansion of $\exp (-S)$ at third order in $S$, which cancels with the factor of 6 coming from the different products of the terms in $S^3$. Also, we have to take into account a minus sign coming from the expansion of $\exp (-S)$ at third order in $S$. Therefore, using the results
of the integrals summarized in the appendix, we obtain
\be
 D_7= 4i\alpha^4 
\left|\displaystyle\frac{k^2}{\mu^2} \right|^{-\epsilon}
\left[\displaystyle\frac{1}{\bar{k}} \left(J^{[\underline{ab}]} + \displaystyle\frac{(N^{ab} - N^{a'b'})}{2} \right) + \displaystyle\frac{k}{\bar k^2} \left(  {\bar J}^{[\underline{ab}]} + \displaystyle\frac{({\hat N}^{ab} - {\hat N}^{a'b'})}{2} \right)\right].
\ee 

\section{One Loop OPE with One Classical Field} 

In this section, we are going to present the results of the OPE's to each expectation value that appears in (\ref{J_0J_0}), specifying which diagrams contribute to each one. But first note that 

\begin{equation}
\langle J^{\au} (y) J^{\bu} (z) \rangle = \langle J^a (y) J^b (z) \rangle + \langle J^{a'} (y) J^{b'} (z) \rangle .
\end{equation}

\noindent So, due to the different contributions to each type of index $a$ and $a'$, the results for $\langle J^a (y) J^b (z) \rangle$ and $\langle J^{a'} (y) J^{b'} (z) \rangle$  will be presented separately. We will use the notation $I_n$ for the contribution of the  n-th diagram and we are going to write the contributions for each expectation value in (\ref {expJJ}). The summary of the results is presented in subsection C.

\subsection{Results for $\langle J^a (y) J^b (z) \rangle$}

To the expectation value $\alpha^2 \langle \partial X^{a} (y) \partial X^{b} (z) \rangle$, the diagrams 4, 5, 6, 7, 8 and 21 contribute:

\begin{eqnarray}
 I_4 = 4i\alpha^4 ({\bar J}^{[ab]} - \frac{{\hat N}^{ab}}{2}) \frac{k}{{\bar k}^2} \left|\frac{k^2}{\mu^2}\right|^{-\epsilon} \left(\frac{1}{\epsilon} + 2\right) 
+4i\alpha^4 (J^{[ab]} - \frac{N^{ab}}{2}) \frac{1}{{\bar k}} \left|\frac{k^2}{\mu^2}\right|^{-\epsilon} \left(\frac{1}{\epsilon} + 2\right) .\nonumber \\
I_5 = -2i\alpha^4 {\bar J}^{[ab]} \frac{k}{{\bar k}^2} \left|\frac{k^2}{\mu^2}\right|^{-\epsilon} \left(\frac{1}{\epsilon} + 2\right) 
-2i\alpha^4 J^{[ab]} \frac{1}{{\bar k}} \left|\frac{k^2}{\mu^2}\right|^{-\epsilon} \left(\frac{1}{\epsilon} + 2\right). \nonumber\\
I_6 = -2i\alpha^4 {\bar J}^{[ab]} \frac{k}{{\bar k}^2} \left|\frac{k^2}{\mu^2}\right|^{-\epsilon} \left(\frac{1}{\epsilon} + 2\right) 
-2i\alpha^4 J^{[ab]} \frac{1}{{\bar k}} \left|\frac{k^2}{\mu^2}\right|^{-\epsilon} \left(\frac{1}{\epsilon} + 2\right) .\nonumber\\
I_7 = 2i\alpha^4 (2{\bar J}^{[ab]} + {\hat N}^{ab}) \frac{k}{{\bar k}^2} \left|\frac{k^2}{\mu^2}\right|^{-\epsilon} 
+2i\alpha^4 (2J^{[ab]} + N^{ab}) \frac{1}{{\bar k}} \left|\frac{k^2}{\mu^2}\right|^{-\epsilon} . \nonumber\\
 I_8 = 4i\alpha^4 ({\bar J}^{[ab]} - \frac{{\hat N}^{ab}}{2}) \frac{k}{{\bar k}^2} \left|\frac{k^2}{\mu^2}\right|^{-\epsilon} \left(\frac{1}{\epsilon} + 2\right) 
+4i\alpha^4 (J^{[ab]} - \frac{N^{ab}}{2}) \frac{1}{{\bar k}} \left|\frac{k^2}{\mu^2}\right|^{-\epsilon} \left(\frac{1}{\epsilon} + 2\right) .\nonumber\\
I_{21} = \frac{3i\alpha^4}{4} {\hat N}^{ab} \frac{k}{{\bar k}^2} \left|\frac{k^2}{\mu^2}\right|^{-\epsilon} \left(-\frac{4}{\epsilon} - \frac{7}{2}\right) +\frac{3i\alpha^4}{4} N^{ab} \frac{1}{{\bar k}} \left|\frac{k^2}{\mu^2}\right|^{-\epsilon} \left(-\frac{4}{\epsilon} - \frac{7}{2}\right),
\end{eqnarray}

\noindent giving as result
\begin{eqnarray}
\alpha^2  \langle \partial X^{a} (y) \partial X^{b} (z) \rangle = i\alpha^4 \frac{k}{{\bar k}^2} \left|\frac{k^2}{\mu^2}\right|^{-\epsilon} \left( 4 {\bar J}^{[ab]} \left(\frac{1}{\epsilon} + 3\right) - {\hat N}^{ab} \left( \frac{7}{\epsilon} + \frac{69}{8} \right) \right) \nonumber\\
+ i\alpha^4 \frac{1}{{\bar k}} \left|\frac{k^2}{\mu^2}\right|^{-\epsilon} \left( 4 J^{[ab]}\left(\frac{1}{\epsilon} + 3\right) - N^{ab} \left( \frac{7}{\epsilon} + \frac{69}{8} \right) \right).
\end{eqnarray}

To the expectation value $-\alpha^3 \langle \partial X^{a} (y) \partial X^{\alpha} X^{\beta} (z) \rangle \gamma^{b}_{\alpha\beta}$, the diagrams 2, 3 and 19 contribute:

\begin{eqnarray}
I_2  = -2i\alpha^4 ({\bar J}^{[ab]} + {\hat N}^{ab}) \frac{k}{{\bar k}^2} \left|\frac{k^2}{\mu^2}\right|^{-\epsilon} \left(\frac{1}{\epsilon} + 2\right) -2i\alpha^4 (J^{[ab]} - N^{ab}) \frac{1}{{\bar k}} \left|\frac{k^2}{\mu^2}\right|^{-\epsilon} \left(\frac{1}{\epsilon} + 2\right). \nonumber\\
I_3 = i\alpha^4 \left({\bar J}^{[ab]} +\frac{{\hat N}^{ab}}{3!} \right) \frac{k}{{\bar k}^2} \left|\frac{k^2}{\mu^2}\right|^{-\epsilon} \left(\frac{1}{\epsilon} + 2\right) +i\alpha^4 \left(J^{[ab]} - \frac{N^{ab}}{3!} \right) \frac{1}{{\bar k}} \left|\frac{k^2}{\mu^2}\right|^{-\epsilon} \left(\frac{1}{\epsilon} + 2\right). \nonumber\\
I_{19}  = -i\alpha^4 \left({\bar J}^{[ab]} + \frac{{\hat N}^{ab}}{2} \right)\frac{k}{{\bar k}^2} \left|\frac{k^2}{\mu^2}\right|^{-\epsilon} \left(\frac{1}{\epsilon} + 4\right) + i\alpha^4 \frac{1}{\bar k} \left(J^{[ab]} + \frac{N^{ab}}{2}\right) \left|\frac{k^2}{\mu^2}\right|^{-\epsilon} \frac{1}{\epsilon},
\end{eqnarray}

\noindent giving as result

\begin{eqnarray}
-\alpha^3 \langle \partial X^{a} (y) \partial X^{\alpha} X^{\beta} (z) \rangle \gamma^{b}_{\alpha\beta} &=& -2i\alpha^4 {\bar J}^{[ab]} \frac{k}{{\bar k}^2} \left|\frac{k^2}{\mu^2}\right|^{-\epsilon} \left(\frac{1}{\epsilon} + 3 \right) - i\alpha^4 \frac{{\hat N}^{ab}}{3} \frac{k}{{\bar k}^2} \left|\frac{k^2}{\mu^2}\right|^{-\epsilon} \left(\frac{7}{\epsilon} + 17\right) \nonumber\\
&-& 2i\alpha^4 J^{[ab]} \frac{1}{{\bar k}} \left|\frac{k^2}{\mu^2}\right|^{-\epsilon}  +i\alpha^4 \frac{N^{ab}}{3} \frac{1}{{\bar k}} \left|\frac{k^2}{\mu^2}\right|^{-\epsilon} \left(\frac{7}{\epsilon} + 11\right) .
\end{eqnarray}

To the expectation value $-\alpha^3 \langle \partial X^{a} (y) \partial X^{\hat\alpha} X^{\hat\beta} (z) \rangle \gamma^{b}_{\hat\alpha\hat\beta}$, the diagrams 2, 3 and 19 contribute:

\begin{eqnarray}
I_2 &=& 2i\alpha^4 ({\bar J}^{[ab]} + {\hat N}^{ab}) \frac{k}{{\bar k}^2} \left|\frac{k^2}{\mu^2}\right|^{-\epsilon} \left(\frac{1}{\epsilon} + 2\right) +2i\alpha^4 (J^{[ab]} - N^{ab}) \frac{1}{{\bar k}} \left|\frac{k^2}{\mu^2}\right|^{-\epsilon} \left(\frac{1}{\epsilon} + 2\right). \nonumber\\
I_3 &=& -i\alpha^4 \left({\bar J}^{[ab]} -\frac{{\hat N}^{ab}}{3!} \right) \frac{k}{{\bar k}^2} \left|\frac{k^2}{\mu^2}\right|^{-\epsilon} \left(\frac{1}{\epsilon} + 2\right) -i\alpha^4 \left(J^{[ab]} + \frac{N^{ab}}{3!} \right) \frac{1}{{\bar k}} \left|\frac{k^2}{\mu^2}\right|^{-\epsilon} \left(\frac{1}{\epsilon} + 2\right) .\nonumber\\
I_{19}  &=& i\alpha^4 \left({\bar J}^{[ab]} + \frac{{\hat N}^{ab}}{2} \right)\frac{k}{{\bar k}^2} \left|\frac{k^2}{\mu^2}\right|^{-\epsilon} \left(\frac{1}{\epsilon} + 4\right) - i\alpha^4 \frac{1}{\bar k} \left(J^{[ab]} + \frac{N^{ab}}{2}\right) \left|\frac{k^2}{\mu^2}\right|^{-\epsilon} \frac{1}{\epsilon},
\end{eqnarray}

\noindent giving as result

\begin{eqnarray}
-\alpha^3 \langle  \partial X^{a} (y) \partial X^{\hat\alpha} X^{\hat\beta} (z) \rangle \gamma^{b}_{\hat\alpha\hat\beta} &=& 2i\alpha^4 {\bar J}^{[ab]} \frac{k}{{\bar k}^2} \left|\frac{k^2}{\mu^2}\right|^{-\epsilon} \left(\frac{1}{\epsilon} + 3 \right) + i\alpha^4 \frac{{\hat N}^{ab}}{3} \frac{k}{{\bar k}^2} \left|\frac{k^2}{\mu^2}\right|^{-\epsilon} \left(\frac{8}{\epsilon} + 19\right) \nonumber\\
&+& 2i\alpha^4 J^{[ab]} \frac{1}{{\bar k}} \left|\frac{k^2}{\mu^2}\right|^{-\epsilon}  -i\alpha^4 \frac{N^{ab}}{3} \frac{1}{{\bar k}} \left|\frac{k^2}{\mu^2}\right|^{-\epsilon} \left(\frac{8}{\epsilon} + 13\right) .
\end{eqnarray}

To the expectation value $\alpha^2 \langle \partial X^{a} (y)  X^e (z) \rangle J^{[cd]} \eta_{e[c} \delta^{b}_{d]}$, only diagram 13 contributes:

\begin{eqnarray}
\alpha^2 \langle \partial X^{a} (y) X^e (z) \rangle J^{[cd]} \eta_{e[c} \delta^{b}_{d]} = I_{13} = -4i\alpha^4 J^{[ab]} \frac{1}{{\bar k}} \left|\frac{k^2}{\mu^2}\right|^{-\epsilon} \left(\frac{1}{\epsilon} +2\right) .
\end{eqnarray}

To the expectation values $-\displaystyle\frac{\alpha^3}{2} \langle \partial X^{a} (y)  X^{\gamma} X^{\beta} (z) \rangle J^{[cd]} (\gamma_{cd})_{\gamma}^{\:\:\:\alpha} \gamma^{b}_{\alpha\beta}$ and

$-\displaystyle\frac{\alpha^3}{2} \langle \partial X^{a} (y) X^{\hat\gamma} X^{\hat\beta} (z) \rangle J^{[cd]}(\gamma_{cd})_{\hat\gamma}^{\:\:\: \hat\alpha} \gamma^{b}_{\hat\alpha\hat\beta}$  only diagram 15 contributes; however, it cancels:

\begin{eqnarray}
-\displaystyle\frac{\alpha^3}{2} \langle \partial X^{a} (y)  X^{\gamma} X^{\beta} (z)\rangle J^{[cd]} (\gamma_{cd})_{\gamma}^{\:\:\:\alpha} \gamma^{b}_{\alpha\beta} = I_{15} = 0, \nonumber\\
-\displaystyle\frac{\alpha^3}{2} \langle \partial X^{a} (y)  X^{\hat\gamma} X^{\hat\beta} (z)\rangle J^{[cd]} (\gamma_{cd})_{\hat\gamma}^{\:\:\: \hat\alpha} \gamma^{b}_{\hat\alpha\hat\beta} = I_{15} = 0 .
\end{eqnarray}

To the expectation value $-\alpha^3 \langle\partial X^{\alpha} X^{\beta} (y) \partial X^b (z)\rangle \gamma^{a}_{\alpha\beta}$, the diagrams 1, 10 and 14 contribute:

\begin{eqnarray}
I_1 = i\alpha^4 \left({\bar J}^{[ab]} +\frac{{\hat N}^{ab}}{3!} \right) \frac{k}{{\bar k}^2} \left|\frac{k^2}{\mu^2}\right|^{-\epsilon} \left(\frac{1}{\epsilon} + 2\right) +i\alpha^4 \left(J^{[ab]} - \frac{N^{ab}}{3!} \right) \frac{1}{{\bar k}} \left|\frac{k^2}{\mu^2}\right|^{-\epsilon} \left(\frac{1}{\epsilon} + 2\right). \nonumber\\
I_{10} = -2i\alpha^4 ({\bar J}^{[ab]} + {\hat N}^{ab}) \frac{k}{{\bar k}^2} \left|\frac{k^2}{\mu^2}\right|^{-\epsilon} \left(\frac{1}{\epsilon} + 2\right) -2i\alpha^4 (J^{[ab]} - N^{ab}) \frac{1}{{\bar k}} \left|\frac{k^2}{\mu^2}\right|^{-\epsilon} \left(\frac{1}{\epsilon} + 2\right). \nonumber\\
I_{14} = -i\alpha^4 \left({\bar J}^{[ab]} + \frac{{\hat N}^{ab}}{2} \right)\frac{k}{{\bar k}^2} \left|\frac{k^2}{\mu^2}\right|^{-\epsilon} \left(\frac{1}{\epsilon} + 4\right) + i\alpha^4 \frac{1}{\bar k} \left(J^{[ab]} + \frac{N^{ab}}{2}\right) \left|\frac{k^2}{\mu^2}\right|^{-\epsilon} \frac{1}{\epsilon},
\end{eqnarray}

\noindent giving as result

\begin{eqnarray}
-\alpha^3 \langle \partial X^{\alpha} X^{\beta} (y) \partial X^b (z)\rangle \gamma^{a}_{\alpha\beta} &=& -2i\alpha^4 {\bar J}^{[ab]} \frac{k}{{\bar k}^2} \left|\frac{k^2}{\mu^2}\right|^{-\epsilon} \left(\frac{1}{\epsilon} + 3 \right) - i\alpha^4 \frac{{\hat N}^{ab}}{3} \frac{k}{{\bar k}^2} \left|\frac{k^2}{\mu^2}\right|^{-\epsilon} \left(\frac{7}{\epsilon} + 17\right) \nonumber\\
&-& 2i\alpha^4 J^{[ab]} \frac{1}{{\bar k}} \left|\frac{k^2}{\mu^2}\right|^{-\epsilon}  +i\alpha^4 \frac{N^{ab}}{3} \frac{1}{{\bar k}} \left|\frac{k^2}{\mu^2}\right|^{-\epsilon} \left(\frac{7}{\epsilon} + 11\right) .
\end{eqnarray}

To the expectation value $-\alpha^3 \langle\partial X^{\hat\alpha} X^{\hat\beta} (y) \partial X^b (z)\rangle \gamma^{a}_{\hat\alpha\hat\beta}$, the diagrams 1, 10 and 14 contribute:

\begin{eqnarray}
I_1 &=& -i\alpha^4 \left({\bar J}^{[ab]} -\frac{{\hat N}^{ab}}{3!} \right) \frac{k}{{\bar k}^2} \left|\frac{k^2}{\mu^2}\right|^{-\epsilon} \left(\frac{1}{\epsilon} + 2\right) -i\alpha^4 \left(J^{[ab]} + \frac{N^{ab}}{3!} \right) \frac{1}{{\bar k}} \left|\frac{k^2}{\mu^2}\right|^{-\epsilon} \left(\frac{1}{\epsilon} + 2\right). \nonumber\\
I_{10} &=& 2i\alpha^4 ({\bar J}^{[ab]} + {\hat N}^{ab}) \frac{k}{{\bar k}^2} \left|\frac{k^2}{\mu^2}\right|^{-\epsilon} \left(\frac{1}{\epsilon} + 2\right) +2i\alpha^4 (J^{[ab]} - N^{ab}) \frac{1}{{\bar k}} \left|\frac{k^2}{\mu^2}\right|^{-\epsilon} \left(\frac{1}{\epsilon} + 2\right) .\nonumber\\
I_{14}  &=& i\alpha^4 \left({\bar J}^{[ab]} + \frac{{\hat N}^{ab}}{2} \right)\frac{k}{{\bar k}^2} \left|\frac{k^2}{\mu^2}\right|^{-\epsilon} \left(\frac{1}{\epsilon} + 4\right) - i\alpha^4 \frac{1}{\bar k} \left(J^{[ab]} + \frac{N^{ab}}{2}\right) \left|\frac{k^2}{\mu^2}\right|^{-\epsilon} \frac{1}{\epsilon},
\end{eqnarray}

\noindent giving as result

\begin{eqnarray}
-\alpha^3 \langle\partial X^{\hat\alpha} X^{\hat\beta} (y) \partial X^b (z)\rangle \gamma^{a}_{\hat\alpha\hat\beta} &=& 2i\alpha^4 {\bar J}^{[ab]} \frac{k}{{\bar k}^2} \left|\frac{k^2}{\mu^2}\right|^{-\epsilon} \left(\frac{1}{\epsilon} + 3 \right) + i\alpha^4 \frac{{\hat N}^{ab}}{3} \frac{k}{{\bar k}^2} \left|\frac{k^2}{\mu^2}\right|^{-\epsilon} \left(\frac{8}{\epsilon} + 19\right) \nonumber\\
&+& 2i\alpha^4 J^{[ab]} \frac{1}{{\bar k}} \left|\frac{k^2}{\mu^2}\right|^{-\epsilon}  -i\alpha^4 \frac{N^{ab}}{3} \frac{1}{{\bar k}} \left|\frac{k^2}{\mu^2}\right|^{-\epsilon} \left(\frac{8}{\epsilon} + 13\right) .
\end{eqnarray}

To the expectation value $\alpha^4 \langle\partial X^{\alpha} X^{\beta} (y) \partial X^{\hat\gamma} X^{\hat\delta} (z) \rangle \gamma^{a}_{\alpha\beta} \gamma^{b}_{\hat\gamma\hat\delta}$ only diagram 9 contributes:

\begin{eqnarray}
\alpha^4 \langle\partial X^{\alpha} X^{\beta} (y) \partial X^{\hat\gamma} X^{\hat\delta} (z) \rangle \gamma^{a}_{\alpha\beta} \gamma^{b}_{\hat\gamma\hat\delta} &=& I_9 = -2i\alpha^4 \left({\bar J}^{[ab]} + \frac{{\hat N}^{ab}}{2} \right) \frac{k}{{\bar k}^2} \left|\frac{k^2}{\mu^2}\right|^{-\epsilon} \left(\frac{1}{\epsilon} + 3 \right) \nonumber\\
&+& 2i\alpha^4 \left(J^{[ab]} + \frac{N^{ab}}{2} \right) \frac{1}{{\bar k}} \left|\frac{k^2}{\mu^2}\right|^{-\epsilon} \left(\frac{1}{\epsilon} + 1 \right) .
\end{eqnarray}

To the expectation value $\alpha^4 \langle \partial X^{\hat\alpha} X^{\hat\beta} (y) \partial X^{\gamma} X^{\delta} (z) \rangle \gamma^{a}_{\hat\alpha\hat\beta} \gamma^{b}_{\gamma\delta}$ only diagram 9 contributes:

\begin{eqnarray}
\alpha^4 \langle \partial X^{\hat\alpha} X^{\hat\beta} (y) \partial X^{\gamma} X^{\delta} (z) \rangle \gamma^{a}_{\hat\alpha\hat\beta} \gamma^{b}_{\gamma\delta} &=& I_9 = -2i\alpha^4 \left({\bar J}^{[ab]} + \frac{{\hat N}^{ab}}{2} \right) \frac{k}{{\bar k}^2} \left|\frac{k^2}{\mu^2}\right|^{-\epsilon} \left(\frac{1}{\epsilon} + 3 \right) \nonumber\\
&+& 2i\alpha^4 \left(J^{[ab]} + \frac{N^{ab}}{2} \right) \frac{1}{{\bar k}} \left|\frac{k^2}{\mu^2}\right|^{-\epsilon} \left(\frac{1}{\epsilon} + 1 \right) .
\end{eqnarray}

To the expectation value $-\alpha^3 \langle \partial X^{\alpha} X^{\beta} (y)  X^{e} (z) \rangle \gamma^{a}_{\alpha\beta} J^{[cd]} \eta_{e[c} \delta^{b}_{d]}$, only diagram 20 contributes:

\begin{eqnarray}
-\alpha^3 \langle \partial X^{\alpha} X^{\beta} (y)  X^{e} (z) \rangle \gamma^{a}_{\alpha\beta} J^{[cd]} \eta_{e[c} \delta^{b}_{d]} = I_{20} = 2i\alpha^4 J^{[ab]} \frac{1}{{\bar k}} \left|\frac{k^2}{\mu^2}\right|^{-\epsilon} \left(\frac{1}{\epsilon} + 2 \right) .
\end{eqnarray}

To the expectation value $-\alpha^3 \langle \partial X^{\hat\alpha} X^{\hat\beta} (y)  X^{e} (z) \rangle \gamma^{a}_{\hat\alpha\hat\beta} J^{[cd]} \eta_{e[c} \delta^{b}_{d]}$, only diagram 20 contributes:

\begin{eqnarray}
-\alpha^3 \langle \partial X^{\hat\alpha} X^{\hat\beta} (y)  X^{e} (z) \rangle \gamma^{a}_{\hat\alpha\hat\beta} J^{[cd]} \eta_{e[c} \delta^{b}_{d]} = I_{20} = -2i\alpha^4 J^{[ab]} \frac{1}{{\bar k}} \left|\frac{k^2}{\mu^2}\right|^{-\epsilon} \left(\frac{1}{\epsilon} + 2 \right) .
\end{eqnarray}

To the expectation value $\displaystyle\frac{\alpha^4}{2} \langle \partial X^{\alpha} X^{\beta} (y)  X^{\hat\gamma} X^{\hat\delta}(z)\rangle \gamma^{a}_{\alpha\beta} J^{[cd]} (\gamma_{cd})_{\hat\gamma}^{\:\:\:\hat\lambda} \gamma^{b}_{\hat\lambda\hat\delta}$ only diagram 16 contributes; however, it cancels:

\begin{eqnarray}
\displaystyle\frac{\alpha^4}{2} \langle \partial X^{\alpha} X^{\beta} (y)  X^{\hat\gamma} X^{\hat\delta}(z)\rangle \gamma^{a}_{\alpha\beta} J^{[cd]} (\gamma_{cd})_{\hat\gamma}^{\:\:\:\hat\lambda} \gamma^{b}_{\hat\lambda\hat\delta} = I_{16} = 0.
\end{eqnarray}

To the expectation value $\displaystyle\frac{\alpha^4}{2} \langle \partial X^{\hat\alpha} X^{\hat\beta} (y)  X^{\gamma} X^{\delta}(z) \rangle \gamma^{a}_{\hat\alpha\hat\beta} J^{[cd]} (\gamma_{cd})_{\gamma}^{\:\:\:\lambda} \gamma^{b}_{\lambda\delta}$ only diagram 16 contributes; however, it cancels:

\begin{eqnarray}
\displaystyle\frac{\alpha^4}{2} \langle \partial X^{\hat\alpha} X^{\hat\beta} (y)  X^{\gamma} X^{\delta}(z)\rangle \gamma^{a}_{\hat\alpha\hat\beta} J^{[cd]} (\gamma_{cd})_{\gamma}^{\:\:\:\lambda} \gamma^{b}_{\lambda\delta} = I_{16} = 0.
\end{eqnarray}

 To the expectation value $\alpha^2 \langle  X^e (y) \partial X^{b} (z) \rangle J^{[cd]} \eta_{e[c} \delta^{a}_{d]}$, only diagram 12 contributes:

\begin{eqnarray}
\alpha^2 \langle  X^e (y) \partial X^{b} (z) \rangle J^{[cd]} \eta_{e[c} \delta^{a}_{d]} = I_{12} = -4i\alpha^4 J^{[ab]} \frac{1}{{\bar k}} \left|\frac{k^2}{\mu^2}\right|^{-\epsilon} \left(\frac{1}{\epsilon} +2\right) .
\end{eqnarray}

To the expectation value $-\alpha^3 \langle   X^{e} (y) \partial X^{\alpha} X^{\beta} (z) \rangle \gamma^{b}_{\alpha\beta} J^{[cd]} \eta_{e[c} \delta^{a}_{d]}$, only diagram 18 contributes:

\begin{eqnarray}
-\alpha^3 \langle   X^{e} (y) \partial X^{\alpha} X^{\beta} (z) \rangle \gamma^{b}_{\alpha\beta} J^{[cd]} \eta_{e[c} \delta^{a}_{d]} = I_{18} = 2i\alpha^4 J^{[ab]} \frac{1}{{\bar k}} \left|\frac{k^2}{\mu^2}\right|^{-\epsilon} \left(\frac{1}{\epsilon} + 2 \right) .
\end{eqnarray}

To the expectation value $-\alpha^3 \langle   X^{e} (y) \partial X^{\hat\alpha} X^{\hat\beta} (z) \rangle \gamma^{b}_{\hat\alpha\hat\beta} J^{[cd]} \eta_{e[c} \delta^{a}_{d]}$, only diagram 18 contributes:

\begin{eqnarray}
-\alpha^3 \langle   X^{e} (y) \partial X^{\hat\alpha} X^{\hat\beta} (z)\rangle \gamma^{b}_{\hat\alpha\hat\beta} J^{[cd]} \eta_{e[c} \delta^{a}_{d]} = I_{18} = -2i\alpha^4 J^{[ab]} \frac{1}{{\bar k}} \left|\frac{k^2}{\mu^2}\right|^{-\epsilon} \left(\frac{1}{\epsilon} + 2 \right) .
\end{eqnarray}

To the expectation values $-\displaystyle\frac{\alpha^3}{2} \langle  X^{\gamma} X^{\beta} (y) \partial X^{b} (z)\rangle J^{[cd]} (\gamma_{cd})_{\gamma}^{\:\:\:\alpha} \gamma^{a}_{\alpha\beta}$ and

$-\displaystyle\frac{\alpha^3}{2} \langle  X^{\hat\gamma} X^{\hat\beta} (y) \partial X^{b} (z)\rangle J^{[cd]} (\gamma_{cd})_{\hat\gamma}^{\:\:\: \hat\alpha} \gamma^{a}_{\hat\alpha\hat\beta}$ only diagram 11 contributes; however, it cancels for the two cases:

\begin{eqnarray}
-\displaystyle\frac{\alpha^3}{2} \langle  X^{\gamma} X^{\beta} (y) \partial X^{b} (z)\rangle J^{[cd]} (\gamma_{cd})_{\gamma}^{\:\:\:\alpha} \gamma^{a}_{\alpha\beta} = I_{11} = 0 \\
-\displaystyle\frac{\alpha^3}{2} \langle  X^{\hat\gamma} X^{\hat\beta} (y) \partial X^{b} (z)\rangle J^{[cd]} (\gamma_{cd})_{\hat\gamma}^{\:\:\: \hat\alpha} \gamma^{a}_{\hat\alpha\hat\beta} = I_{11} = 0 .
\end{eqnarray}

To the expectation value $\displaystyle\frac{\alpha^4}{2} \langle X^{\hat\gamma} X^{\hat\delta}(y) \partial X^{\alpha} X^{\beta} (z)\rangle \gamma^{b}_{\alpha\beta}  J^{[cd]} (\gamma_{cd})_{\hat\gamma}^{\:\:\:\hat\lambda} \gamma^{a}_{\hat\lambda\hat\delta}$ only diagram 17 contributes; however, it cancels:

\begin{eqnarray}
\displaystyle\frac{\alpha^4}{2}\langle   X^{\hat\gamma} X^{\hat\delta}(y) \partial X^{\alpha} X^{\beta} (z)\rangle \gamma^{b}_{\alpha\beta} J^{[cd]} (\gamma_{cd})_{\hat\gamma}^{\:\:\:\hat\lambda} \gamma^{a}_{\hat\lambda\hat\delta} = I_{17} = 0.
\end{eqnarray}

To the expectation value $\displaystyle\frac{\alpha^4}{2} \langle  X^{\gamma} X^{\delta}(y) \partial X^{\hat\alpha} X^{\hat\beta} (z) \rangle \gamma^{b}_{\hat\alpha\hat\beta} J^{[cd]} (\gamma_{cd})_{\gamma}^{\:\:\:\lambda} \gamma^{a}_{\lambda\delta}$ only diagram 17 contributes; however, it cancels:

\begin{eqnarray}
\displaystyle\frac{\alpha^4}{2} \langle   X^{\gamma} X^{\delta}(y) \partial X^{\hat\alpha} X^{\hat\beta} (z) \rangle \gamma^{b}_{\hat\alpha\hat\beta} J^{[cd]} (\gamma_{cd})_{\gamma}^{\:\:\:\lambda} \gamma^{a}_{\lambda\delta} = I_{17} = 0.
\end{eqnarray}

\subsection{Results for $\langle J^{a'} (y) J^{b'} (z) \rangle$}

The results for the expectation value $\langle J^{a'} (y) J^{b'} (z) \rangle$ are analogous to the results of subsection A. We just need to be careful with the signals of some vertices. The result is:
\begin{equation}
\langle J^{a'} (y) J^{b'} (z) \rangle =  -i\alpha^4 \frac{N^{a'b'}}{\bar k} \left|\frac{k^2}{\mu^2} \right|^{-\epsilon} \left( \frac{1}{3\epsilon} -\frac{65}{24} \right) +i\alpha^4 {\hat N}^{a'b'} \frac{k}{{\bar k}^2} \left|\frac{k^2}{\mu^2} \right|^{-\epsilon} \left( \frac{7}{3\epsilon} +\frac{193}{24} \right) .
\end{equation}

\subsection{Summary of the results}

Finally, summing up all the expectation values presented in subsections A and B, we have

\begin{eqnarray}
J^{a'} (x) J^{b'} (y) \rightarrow -i\alpha^4 \frac{N^{a'b'}}{\bar k} \left|\frac{k^2}{\mu^2} \right|^{-\epsilon} \left( \frac{1}{3\epsilon} -\frac{65}{24} \right) +i\alpha^4 {\hat N}^{a'b'} \frac{k}{{\bar k}^2} \left|\frac{k^2}{\mu^2} \right|^{-\epsilon} \left( \frac{7}{3\epsilon} +\frac{193}{24} \right) \\
J^a (x) J^b (y) \rightarrow -i\alpha^4 \frac{N^{ab}}{\bar k} \left|\frac{k^2}{\mu^2} \right|^{-\epsilon} \left( \frac{17}{3\epsilon} +\frac{191}{24} \right) -i\alpha^4 {\hat N}^{ab} \frac{k}{{\bar k}^2} \left|\frac{k^2}{\mu^2} \right|^{-\epsilon} \left( \frac{25}{3\epsilon} +\frac{319}{24} \right) .
\end{eqnarray}

The results can now be expressed in coordinate space
using the following:

\begin{eqnarray}
\frac{i k}{\bar{k}^2} \Rightarrow \frac{\bar{z}-\bar{w}}{\left(z-w\right)^2} . \label{kkbarra}\\
\frac{-i}{\bar{k}} \Rightarrow \frac{1}{z-w}  \label{kkbarraquad}. \\
\frac{-i}{\bar{k}}\left[\frac{1}{\epsilon}-\ln|\frac{k}{\mu}|^2\right] \Rightarrow \ln|z-w|^2\frac{1}{z-w}.\label{kbarraepsilon}\\
\frac{-ik}{\bar{k}^2}\left[\frac{1}{\epsilon}- \ln|\frac{k}{\mu}|^2+ 2\right] \Rightarrow \ln|z-w|^2\frac{({\bar z}-{\bar w})}{(z-w)^2}. \label{kbarraquadepsilon}
\end{eqnarray}

The terms (\ref{kkbarra}) and (\ref{kkbarraquad})  can be derived by simple Inverse Fourier transform. For (\ref{kbarraepsilon}) we calculate the expectation value  $\langle \partial X(z) X(z) X(w) X(w) \rangle$ in momentum and coordinate space, and then compare them; the same is done for (\ref{kbarraquadepsilon}), calculating $\langle \partial X (z) X(z)\int d^2 u \partial X(u) X(u) \partial X(w)X(w) \rangle$ in momentum and in coordinate space. The results in coordinate space are
\begin{eqnarray}
J^{a'} (x) J^{b'} (y) 
& \rightarrow &\frac{N^{a'b'}}{(x-y)} \left(\frac{1}{3} \ln |x-y|^2 - \frac{65}{24} \right)  + {\hat N}^{a'b'} \frac{({\bar x}- {\bar y})}{(x-y)^2} \left(-\frac{7}{3} \ln |x-y|^2 + \frac{81}{24} \right) .\\
J^a (x) J^b (y) 
& \rightarrow & \frac{N^{ab}}{(x-y)} \left(\frac{17}{3} \ln |x-y|^2 + \frac{191}{24} \right)  + {\hat N}^{ab} \frac{({\bar x}- {\bar y})}{(x-y)^2} \left(\frac{25}{3} \ln |x-y|^2 + \frac{81}{24} \right).
\end{eqnarray}

Note that the divergences in momentum space OPEs appear as logarithms in coordinate space OPEs, suggesting that the currents Êmay get anomalous dimensions. This is not a surprising result since the $AdS_5 \times S^5$ left-invariant currents are not protected by any symmetry argument. However, as it was shown in reference \cite{quantumopes}, Êwhere it was computed the one loop OPEs between the energy-momentum tensor and the left-invariantÊ currents, the bosonic left-invariant currents do not get anomalous dimensions.ÊÊ On the other hand, the fermionic currents get anomalous dimension contributions. However, Êit was shown that the two types of fermionic currents get the preciseÊ contributions that cancel when combined into the single operator which appears in the energy-momentum tensor. Therefore, the energy-momentum tensor still has zero anomalous dimension. This result suggests that, although the left-invariant currents have logarithmic terms in their one loop OPEs, protected operators constructed with these currents do not have them.

\section{Conclusion}

It was shown in \cite{quantumopes} that, at one loop, there are non trivial cancellations 
in the possible corrections to the double pole of the product of the bosonic currents 
$J^{\underline{a}}(y) J^{\underline{b}}(z)$. Here it was shown that there are contributions to the simple poles. As the simple poles are important in the calculations of the four point function, the results presented here are a step forward in order to get a worldsheet description of pure spinor closed strings in $AdS_5\times S^5$. Since these currents are not holomorphic, there are anti-holomorphic terms in the OPEs. However, if we try to make the same calculation for holomorphic quantities, such as the combination $\lambda^{\alpha}J_{\alpha}$, which is used to define the BRST charge in the $AdS_5\times S^5$, we can see that it is not possible to form diagrams with one classical field and so there is no anti-holomorphic contributions for these quantities. 

As the tree level calculation of reference \cite{Puletti:2006vb},  the one loop  OPEs calculated here respect the Z4-grading of the $psu(2, 2|4)$ super-algebra, reflecting the fact that all the interactions  respect the Z4-automorphism of the super-algebra. Important to note that all terms proportional to the currents $J^{\au\bu}$, $\bar{J}^{\au\bu}$ cancel and the only contribution comes from the pure spinor Lorentz currents $N^{\au\bu}$ and $\hat{N}^{\au\bu}$. The calculations presented in this work show the way to get the complete one loop current algebra for the pure spinor $AdS_5\times S^5$ string. However, owing to the huge number of vertices coming from background expansion, we hope to develop some algebraic computational  methods in order to complete such a task.

As a final remark, we would like to draw attention to the reference \cite{benichou}, where the author presents another method to compute the current algebra. In this paper, this method was used to compute the current algebra at tree level. However, the author states that it can be generalized to compute quantum corrections to the algebra of the currents. It is a good task to investigate in a future work.

\section*{Acknowledgments}

D\'afni F. Z. Marchioro would like to thank ICTP for financial support. Daniel L. Nedel would like to thank CNPq for financial support. 

This paper is dedicated to Yasmim Marchioro Nedel. 

\section{Appendix}

\subsection{Diagrams}

\begin{figure}[!h]
\centering
\includegraphics[scale=0.8]{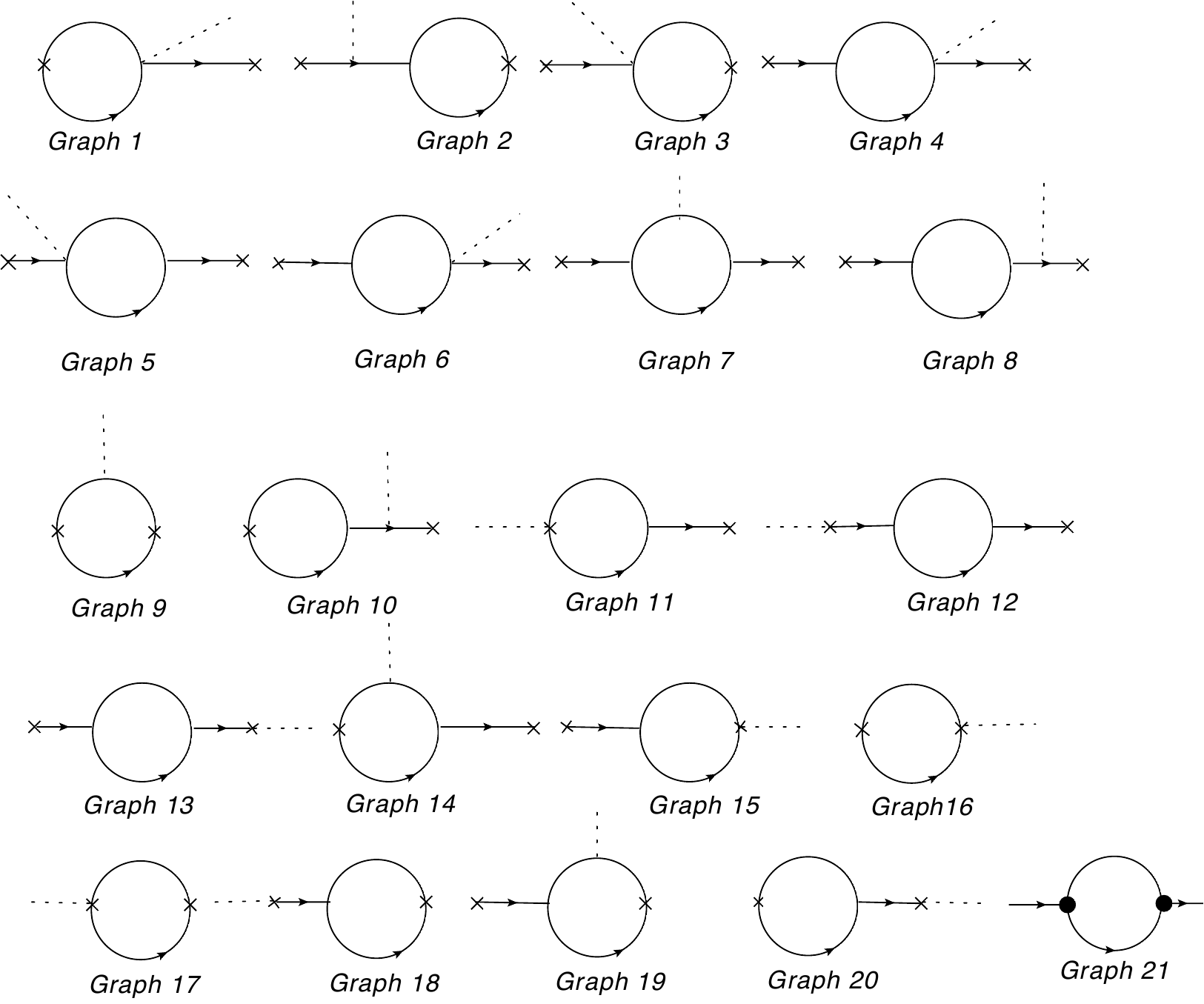}
\caption{Diagrams for one loop contribution. The full lines represent the quantum fields, and the dashed lines represent the classical field. In diagram 21, the dot vertex represents the presence of one quantum pure spinor Lorentz current ($N^{(1)}_{\au\bu}$ or $\hat{N}^{(1)}_{\au\bu}$). }
\end{figure}

\newpage

\subsection{Table of integrals}

\be\label{integralzero} \int d^d m  {{1}\over{|m|^2|m-k|^2}} = - {2\over {k
\bar k}} ({1\over \epsilon} - ln {|k|^2 \over \mu^2})  
\ee

\be\label{integralI}\int d^d m  {{m \bar m}\over{|m|^2|m-k|^2}} = 1 
\ee

\be\label{integralII}\int d^d m {m\over{|m|^2 |m-k|^2}} = -{1\over \pi^\e} 
{1\over \bar k}{{[|k|^2 ]^{-\e}}\over{\mu^{-2\e}}} {{\G (1-\e)^2
\G(\e)}\over{\G(1-2\e)}} = - {1\over{\bar k}} ({1\over \e} - ln{|k|^2
\over{\mu^2}})
\ee

\be\label{integralIII}\int d^d  m {{\bar m}\over{|m|^2 |m-k|^2}} = -{1\over \pi^\e} 
{1\over k}{{[|k|^2 ]^{-\e}}\over{\mu^{-2\e}}} {{\G (1-\e)^2
\G(\e)}\over{\G(1-2\e)}} = - {1\over{ k}} ({1\over \e} - ln{|k|^2
\over{\mu^2}})
\ee

\be\label{integralIV}\int d^d m {m^2\over{|m|^2 |m-k|^2}} = -{1\over \pi^\e} 
{k\over \bar k}{{[|k|^2 ]^{-\e}}\over{\mu^{-2\e}}} {{\G(2-\e)\G (1-\e)
\G(\e)}\over{\G(2-2\e)}} = - {k\over \bar k} ({1\over \e} + 1 -
ln{|k|^2 \over \mu^ 2})
\ee

\be\label{integralV}\int d^d m {{\bar m}^2\over{|m|^2 |m-k|^2}} = -{1\over \pi^\e} 
{\bar k\over k}{{[|k|^2 ]^{-\e}}\over{\mu^{-2\e}}} {{\G(2-\e)\G (1-\e)
\G(\e)}\over{\G(2-2\e)}} = - {\bar k\over k} ({1\over \e} + 1 -
ln{|k|^2 \over \mu^ 2})
\ee

\be\label{integralVI}\int d^d m {m^2 {\bar m}\over{|m|^2 |m-k|^2}} =
{1\over 2\pi^\e} 
k{{[|k|^2 ]^{-\e}}\over{\mu^{-2\e}}} {{\G(1-\e) ^2 \G (1+\e)
}\over{\G(2-2\e)}} = {k\over 2}
\ee

\be\label{integralVII}\int d^d m {m {\bar m}^2\over{|m|^2 |m-k|^2}} =
{1\over 2\pi^\e} 
{\bar k}{{[|k|^2 ]^{-\e}}\over{\mu^{-2\e}}} {{\G(1-\e) ^2 \G (1+\e)
}\over{\G(2-2\e)}} = {{\bar k}\over 2}
\ee

\be\label{integralVIII}\int d^d m {m^2 {\bar m}^2\over{|m|^2 |m-k|^2}} =
{1\over \pi^\e} 
k{\bar k}{{[|k|^2 ]^{-\e}}\over{\mu^{-2\e}}} {{\G(2-\e) ^2 \G (1+\e)
}\over{\G(4-2\e)}} = {{k\bar k}\over 6}
\ee

\be\label{integralIX}\int d^d m {m {\bar m}\over{[|m|^2]^2 |m-k|^2}} = -
{1\over \pi^\e} 
{1\over{k\bar k}}(1+ {2\over \epsilon} - 2 ln {|k|^2 \over \mu^2 } )  
\ee

\be\label{integralX}\int d^d m {m^2 {\bar m}\over{[|m|^2]^2 |m-k|^2}} = -
{1\over \pi^\e} 
{1\over{\bar k}}({1\over \epsilon} -  ln {|k|^2 \over \mu^2 } )  
\ee

\be\label{integralXI}\int d^d m {m^3 {\bar m}\over{[|m|^2]^2 |m-k|^2}} = -
{1\over \pi^\e} 
{k\over{\bar k}}({1\over \epsilon} +1 -  ln {|k|^2 \over \mu^2 } )  
\ee

\be\label{integralXII}\int d^d m {m^2 {\bar m^2}\over{[|m|^2]^2
|m-k|^2}} = {3\over 2}
\ee

\end{document}